\documentclass[aps,prl,groupedaddress,a4paper,twocolumn,showpacs]{revtex4}

\usepackage{graphicx}

\begin{document}

\title{Detection of sub-shot-noise spatial correlation in high-gain parametric down-conversion}

\author{O. Jedrkiewicz$^1$, Y.-K Jiang$^2$, E. Brambilla$^1$, A.
Gatti$^1$, M. Bache$^1$, L. A. Lugiato$^1$, and P. Di Trapani$^1$}

\affiliation{$^1$ INFM, Dipartimento di Fisica e Matematica,
Universita' dell'Insubria, Via Valleggio 11, 22100 Como, Italy\\
$^2$ Department of Electronic Science and Applied Physics, Fuzhou
University, 350002 Fuzhou, China}

\date{\today}

\begin{abstract}
Using a 1GW-1ps pump laser pulse in high gain parametric
down-conversion allows us to detect sub-shot-noise spatial quantum
correlation with up to one hundred photoelectrons per mode, by
means of a high efficiency CCD. The statistics is performed in
single-shot over independent spatial replica of the system. The
paper highlights the evidence of quantum correlation between
symmetrical signal and idler spatial areas in the far field, in
the high gain regime. In accordance with the predictions of
numerical calculations the observed transition from the quantum to
the classical regime is interpreted as a consequence of the
narrowing of the down-converted beams in the very high gain
regime.

\end{abstract}


\maketitle
\begin{figure}
 \includegraphics[width=8cm]{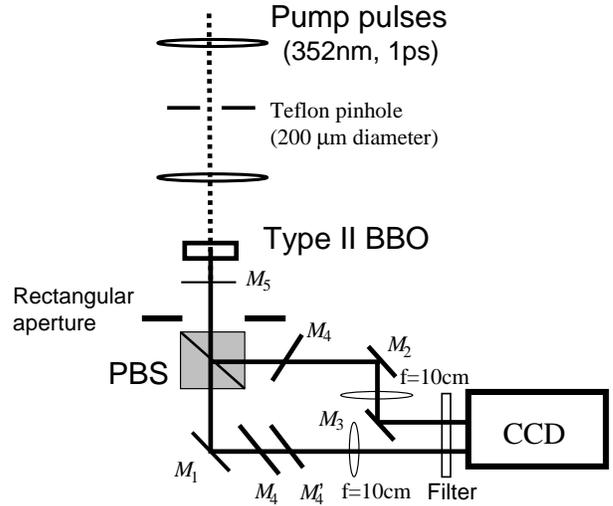}
    \caption{Scheme of the experimental set-up used for the spatial
     correlation measurements (see text).}
    \label{fig1}
   \end{figure}

 Spatial quantum optical fluctuations are studied
because of new potential applications of quantum optical
procedures in parallel processing and multi-channel operation.
Examples are quantum holography \cite{abouraddy01}, the quantum
teleportation of optical images \cite{Sokolov01}, and the
measurements of small displacements beyond the Rayleigh limit
\cite{Treps02}. There is now a large literature on spatial effects
in the spontaneous regime of parametric down-conversion (PDC)
where photons are created one pair at a time \cite{Lugiato02}. The
process of PDC is in fact particularly suitable for the study of
\emph{spatial correlations} because of its large emission
bandwidth in the spatial frequency domain \cite{Devaux00}.
Nevertheless to date most spatial correlation measurements in PDC
have been performed in single photon counts regime \cite{Jost98,
Oemrawsingh02} without evidencing any relevant quantum effects.
The quantum twin beam character of the PDC emission has been
evidenced in \cite{Aytur90} by using low-to-medium pump-power
lasers (${\leq1}$ MW) and relying on statistical ensembles from
different \emph{temporal} replicas of the system. With increasing
gain a transition from the quantum to the classical regime has
been observed \cite{Marable02}. However, recent theoretical
investigations predict multi-mode spatial quantum correlations
(sub-shot-noise photon-number correlation between symmetrical
portions of the signal and idler angular distributions) also for
arbitrarily high gains, provided that the detection area exceeds
the typical size of the mode (coherence area) \cite{Brambilla04,
Gatti99}.
\newline\indent
Here we report on the first quantum spatial measurements of PDC
radiation performed by using a low-repetition rate (2 Hz) pulsed
high-power laser (1GW-1ps). This enables us to tune the PDC to the
high-gain regime while keeping a large pump beam size
(${\sim1mm}$). The huge number of transverse modes (roughly given
by the ratio between: (i) the area of the near-field gain profile
and (ii) the inverse of the angular bandwidth of the PDC process)
allows us to identify regions of the signal and idler beams where
symmetrical signal-idler pixel pairs correspond to independent
spatial replica of the quantum system. We concentrate on a portion
of the parametric fluorescence emitted close to the collinear
direction and within a narrow frequency bandwidth around
degeneracy. The generated pairs of signal and idler
phase-conjugate modes propagate at symmetrical angles with respect
to the pump direction in order to fulfil the phase-matching
constraints, and each pair of symmetrical spots characterizing the
far field represents a spatial replica. Thanks to the very large
number of these, the statistical ensemble averaging necessary for
the quantum measurement can be solely done over the \emph{spatial}
replicas \emph{for each, single, pump-laser pulse}. A
characterization of the system over several shots was only made in
our case in order to verify that the selected spatial replicas are
indeed "statistically" identical, as required for the suitable
definition of the ensemble. The single-shot measurements reveal
sub-shot-noise spatial correlations for a PDC gain $G$ (intensity
amplification factor) leading to the detection of up to $\simeq$
100 photoelectrons per mode. Finally, by numerically solving the
three-waves coupled equations in the framework of a 3D+1 quantum
model, we are able to attribute the observed transition from
quantum to classical regime to the near-field gain narrowing that
occurs in presence of a bell-shaped pump beam, at very-high gain.
\newline\indent
The experimental setup is sketched in Fig.\ \ref{fig1}. The third
harmonic (352 nm) of a 1ps, chirped-pulse amplified Nd:glass laser
(TWINKLE, Light Conversion Ltd.) is used to pump a type II
5x7x4mm${^{3}}$ $\beta$-barium borate (BBO) non-linear crystal,
operated in the regime of parametric amplification of the
vacuum-state fluctuations. The input and output facets of the
crystal are anti-reflection coated at 352 nm and 704 nm,
respectively. The pump beam is spatially filtered and collimated
to a beam waist of approximately 1 mm (FWHM) at the crystal input
facet. The energy of the 352nm pump pulse can be continuously
tuned in the range 0.1-0.4 mJ by means of suitable attenuating
filters and by changing the energy of the 1055nm pump laser pulse,
allowing to have a gain in the range $10 \le G \le 10^3$. The
parametric fluorescence at the horizontally polarized signal and
vertically polarized idler modes is emitted over two cones, whose
apertures depend on the specific wavelengths (see, e.g.,
\cite{Berzanskis99, Rubin96}). The BBO crystal
(${\theta=49.05^\circ}$, ${\phi=0}$) is oriented in order to
generate signal and idler radiation cones tangent in the collinear
direction at the degenerate wavelength
$\omega_s=\omega_i=\omega_p/2$ (\emph{s}, \emph{i} and \emph{p}
referring to signal, idler and pump respectively). The
fluorescence around the collinear direction is selected by means
of a 5mm x 8mm aperture, placed 15 cm from the output facet of the
BBO. The aperture turned out to be necessary in order to prevent
beam clipping by the PBS, otherwise giving rise to substantial
scattered radiation.  The selected portion of the beam is
transmitted through a polarizing beam splitter (PBS), which
separates the signal and idler beams. The latter are finally sent
onto two separate regions of a deep-depletion back illuminated
charged coupled device (CCD) camera \cite {Janesick01} (Roper
Scientific NTE/CCD-400EHRBG1, 16 bits dynamical range, quantum
efficiency $\eta{\approx89 \%}$ at 704 nm at -40${^{\circ}}$C,
dark current and read out noise ${<1 e}$/pixel/s and ${<3
e}$/pixel/s respectively), placed in the common focal plane of the
two lenses (\emph{f}=10 cm) used to image the signal and idler far
fields. The detection array has 1340x400 pixels, with a pixel size
of 20${\mu}$mx20${\mu}$m. Prior to the experiment the CCD was
calibrated with a coherent source allowing the retrieval of
spatial shot-noise statistics in its full dynamic range
\cite{Jiang03}. In our setup the correlation measurements are
performed without using any narrow-band interferential filters
(IFs), in contrast to the case of photon-counting experiments
(coincidence measurements), since IFs unavoidably introduce
relevant transmission losses reducing the visibility of sub-shot
noise correlations. The pump-frequency contribution is removed by
using normal incidence (${M_{5}}$) and at 45 deg (${M_{4}}$)
high-reflectivity (HR) mirrors coated for 352 nm placed before and
after the PBS, respectively, and a low-band pass color filter
(90\% transmission around 704 nm) placed in front of the CCD. A
further HR@352nm mirror ($M'_{4}$) is placed in one of the two
arms at a suitable angle in order to balance the unequal
transmission of the PBS in the two arms. All the optical
components (except the color filter) have antireflection coatings
at 704 nm. The estimated quantum efficiency of each detection
line, which accounts for both the transmission losses and the
detector efficiency, is $\eta_{tot}\simeq$75\%.

\begin{figure}
\includegraphics[width=8cm]{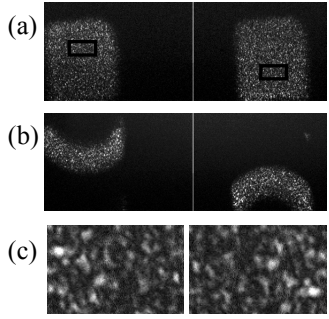}
    \caption{(a) Single-shot far field image recorded by the CCD for
     a pump beam waist ${w_{0}\simeq 1 mm}$ and pump energy
     ${\varepsilon}_{p}\simeq0.3 mJ$. The spatial areas for statistics are delimited by the
   white boxes selected within the degenerate signal and idler modes, spatially localized from the single shot
     image recorded with the 10nm-broad IF (b). (c) Zoom of two symmetrical
     areas of the signal and idler far fields.}
    \label{fig2}
   \end{figure}

Fig.\ \ref{fig2}a shows a typical far field image recorded in a
single shot, where a fairly broadband radiation (\emph{i.e.} the
one transmitted by the rectangular aperture) is acquired in the
signal (left) and idler (right) branches.  The selection of the
desired temporal and angular bandwidth around degeneracy is made
by temporarily inserting in front of the CCD a 10nm wide IF around
704 nm, allowing us to locate the collinear degeneracy point (see
Fig.\ \ref{fig2}b). The data analysis is limited within two
rectangular boxes (black frames in Fig.\ref{fig2}a) corresponding
to an angular bandwidth of 20 mrad x 8 mrad and to a temporal
bandwidth smaller than 10 nm. The selected regions contain 4000
pixels each. In this work we investigate pixel pair correlation,
and since the size of the CCD pixel approximately corresponds to
the physical size of the replica, the ensemble is large enough to
perform the desired statistics. We have observed that much larger
boxes worsen the level of the correlation; this is attributed to
residual scattering owing to diffraction from the borders of the
aperture but also to the contribution of other frequency
components (far from degeneracy) characterized by a lower angular
symmetry between signal and idler cones. A zoom of the selected
areas is presented in Fig.\ref{fig2}c, where the rather
spectacular symmetry of the intensity distribution in the signal
and idler branches shows the twin-beam character of the
phase-conjugate modes.

\begin{figure}
\includegraphics [width=8cm]{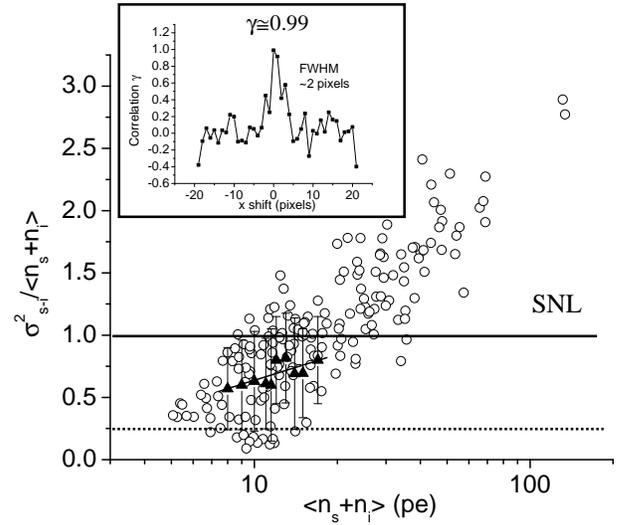}
    \caption{Intensity difference variance ${\sigma^{2}_{s-i}}$ normalized to the SNL
     ${\langle n_{s}+n_{i} \rangle}$. Each point (white circle)
     corresponds to a single shot measurement where the spatial ensemble
     statistics has been performed over a 100 x 40 pixels region.
     The triangles (each one obtained by averaging the experimental
     points corresponding to a certain gain) and their linear fit illustrate the trend
     of the data in the region between   ${\langle n_{s}+n_{i} \rangle}$ =8 and 20.
     Inset: Typical correlation degree profile in the regime where ${\langle n_{s}+n_{i} \rangle\simeq8}$ (see text).}
    \label{fig3}
   \end{figure}

The aim of the experiment is to quantify the symmetrical
\emph{pixel pair} correlation. This is done by measuring the
variance ${\sigma^{2}_{s-i}}$ of the PDC photoelectron-number
difference ${n_{s}-n_{i}}$ of the signal/idler pixel pair versus
the mean total number of down-converted photoelectrons (pe) of the
pixel pair. This variance is ${\sigma^{2}_{s-i}=\langle
(n_{s}-n_{i})^{2}\rangle-\langle n_{s}-n_{i} \rangle^{2}}$ where
the averages are spatial averages performed over all the
symmetrical pixel pairs contained in the chosen regions. Each
single shot of the laser provides a different ensemble,
characterized by its pixel pair average pe number ${\langle
n_{s}+n_{i}\rangle}$, in turn related to the parametric gain. In
the experiment, ensembles corresponding to different gains are
obtained by varying the pump-pulse energy. We note that the
read-out noise of the detector, its dark current, and some
unavoidable light scattered from the pump, signal and idler fields
contribute with a non-negligible background noise to the process.
This is taken into account by applying a standard correction
procedure (see for example \cite{Mosset04}), by subtracting the
background fluctuations ${\sigma^{2}_{b}}$ from the
\emph{effectively measured} variance ${\sigma^{2}_{(s+b)-(i+b)}}$
of the total intensity difference
(signal+background)-(idler+background) obtaining
${\sigma^{2}_{s-i}}={\sigma^{2}_{(s+b)-(i+b)}}-2{\sigma^{2}_{b}}$.
This background noise, having a standard deviation of 7 counts
(${\pm0.1}$ from shot to shot, estimated by repeating the
measurement with the same pump-pulse energy) is measured in
presence of pulse illumination over an area of the same size of
the acquisition area and suitably displaced from the directly
illuminated region. The validity of the data correction procedure
is tested by sending in the setup (with no crystal) through the
PBS a coherent circularly polarized pulsed beam (@704nm), and
verifying for different laser energies that the intensity
difference fluctuations from the two coherent portions of beams
recorded on the CCD lie at the shot noise level.
\newline\indent
Fig.\ \ref{fig3} shows the experimental results where each point
is associated with a different laser shot. The data are normalized
to the shot noise level (SNL), and their statistical spread
accounts for the background correction. Although the noise on the
individual signal and idler beams is found to be very high and
much greater than the standard quantum limit (=${\langle n_{s}
\rangle}$ and ${\langle n_{i} \rangle}$ respectively), we observe
an evident sub-shot noise pixel pair correlation up to gains
characterized by ${\langle n_{s}+n_{i} \rangle{\approx15-18}}$.
Since in that regime the observed transverse size of the coherence
areas (\emph{i.e.} of the modes) is about 2-4 pixels, this
approximately corresponds to 100 pe per mode. We can have an idea
of the transverse size of the mode by looking at the standard
two-dimensional cross-correlation degree ${\gamma=(\langle
n_{s}n_{i} \rangle-\langle n_{s}\rangle \langle
n_{i}\rangle)/\sqrt{\sigma^{2}_{s}\sigma^{2}_{i}}}$, between all
the angularly symmetrical signal and idler pixels contained within
the black boxes (see Fig.\ \ref{fig2}a). This can be plotted for
instance as a function of the horizontal and vertical shifts of
the recorded image on the CCD, keeping fixed the position of the
boxes. In general ${\mid\gamma\mid\leq1}$ with ${\gamma=1}$ for
perfect correlation. A transverse section of the correlation
function obtained from a single-shot image characterized by
${\langle n_{s}+n_{i} \rangle{\approx8}}$ is plotted in the inset
of Fig.\ \ref{fig3} as a function of the horizontal shift $x$ (in
pixels units). As expected, virtually perfect correlation (in our
case the peak value is ${\simeq0.99}$) is obtained for perfect
determination (\emph{i.e.} within one pixel) of the center of
symmetry between the signal and the idler regions.

\begin{figure}
\includegraphics [width=8cm]{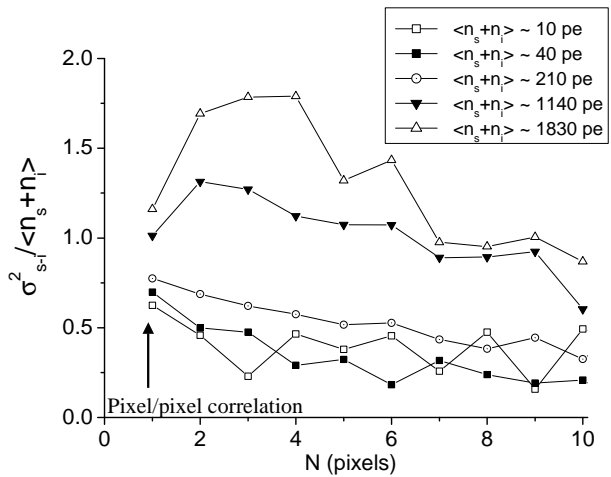}
    \caption{Numerical calculation of ${\sigma^{2}_{s-i}}$ (normalized to SNL)
     between symmetrical portions of signal and idler plotted as a function of the
     detection area represented by N x N binned pixels. Different curves correspond
     to different values of the gain characterized by the mean number of down-converted pe per pixel pair
     ${\langle n_{s}+n_{i} \rangle}$.}
    \label{fig4}
   \end{figure}

In order to interpret the observed transition from quantum to
classical regime we present in Fig.\ \ref{fig4} the results of the
numerical calculations. The full quantum model accounts for the
two transverse and the temporal degrees of freedom with
propagation along the crystal, for the angular and chromatic
material dispersion up to the second order, and for the finite
spatial and temporal widths of the Gaussian pump pulse. The
crystal and input-pulse parameters are those relative to the
experiment. The figure presents ${\sigma^{2}_{s-i}}$, normalized
to the SNL, \emph{vs} the size of the detection area for different
gains. Each point is the result of a statistics performed over one
single laser-shot. The case N=1 corresponds to the experiment. The
simulations (data not shown) outline that, in spite of the fixed
pump-beam diameter, the signal and idler beam diameters at the
crystal output strongly depend on the gain and decrease when the
latter increases. This can be easily interpreted when considering
that the signal and idler beam size maps not the pump-beam profile
but the actual parametric amplification gain profile
${G(\textbf{r})\sim \rm cosh^{2}[{\sigma}A(\textbf{r})L]}$
\cite{Akhmanov} ($L$ being the crystal length, $A$ the pump field
amplitude and ${\sigma}$ a parameter proportional to the setting
characteristics), as long as filtering due to the limited spatial
bandwidth does not take place \cite{Ditrapani98}. On narrowing the
size of the PDC beams, the coherence areas in the far field
(\emph{i.e.} the modes) increase their size, as straightforward
consequence of the convolution theorem in Fourier analysis \cite
{Berzanskis99}. Since revealing quantum correlations requires
detection areas larger (or comparable) to the mode size (as also
discussed in \cite{Brambilla04}), it is necessary when increasing
the gain to have larger detectors in order to obtain
below-shot-noise variance as shown in Fig.\ \ref{fig4}. Note that
Fig.\ \ref{fig4} evidences the transition from quantum to
classical regime in case of single-pixel detection (N=1) for a
gain that is higher than in the experiment. Instead, in the
experiment, excess noise is observed for ${\langle n_{s}+n_{i}
\rangle}>$20, which we attribute first to the effect of residual
scattered light whose contribution grows linearly with the
radiation fluence and is thus expected to overcome the shot noise
at large pumping, and second to the uncertainty in the
determination of the symmetry center of the signal and idler image
portions. In fact simulations have shown that an uncertainty as
small as a few microns (i.e. a fraction of the pixel size,
unavoidable experimentally), prevents to observe sub-shot-noise
correlation as soon as ${\langle{n_s}+{n_i}\rangle}$ exceeds some
tens of pe, while still preserving sub-shot-noise correlation for
smaller gain values. Finally, the maximum level of noise reduction
observed experimentally agrees with the theoretical limit (dotted
line in Fig.\ \ref{fig3}) determined by the total losses of the
system (${\sim1-\eta_{tot}}$ \cite{Brambilla04}), in accordance
with the result of Fig.\ \ref{fig4}.
\newline\indent
In conclusion, we have shown that twin beams of light generated in
parametric down-conversion exhibit sub-shot noise spatial
correlation by measuring an evident quantum noise reduction on the
signal/idler intensity difference. A transition to above
shot-noise correlation is observed as the gain increases. This
quantum-to-classical transition, in agreement with numerical
simulations, is explained as a narrowing of the signal/idler beams
with increased gain. This leads in turn to a larger mode size and
therefore also to the need of larger pixels to observe below
shot-noise correlation \cite{Brambilla04}. This will be the aim of
a future work. To our knowledge, this is the first experimental
investigation of quantum spatial correlations in the high gain
regime, where the huge number of transverse spatial modes is
detected in single shot by means of a high-quantum-efficiency CCD.

This work has been supported by the European Union (QUANTIM
contract IST-2000-26019). M. B. acknowledges support from the
Carlsberg foundation.


\begin{references}

\bibitem{abouraddy01}  A. F. Abouraddy, B. E. A. Saleh, A. V. Sergienko,
and M. C. Teich, Opt. Express \textbf{9}, 498 (2001).

\bibitem{Sokolov01} I. V. Sokolov, M. I. Kolobov, A. Gatti, and L.
A. Lugiato, Opt. Commun. \textbf{193}, 175 (2001).

\bibitem{Treps02} N. Treps, U. Andersen, B. Buchler, P. K. Lam, A.
Maitre, H.-A Bachor, and C. Fabre, Phys. Rev. Lett. \textbf{88},
203601 (2002).

\bibitem{Lugiato02} L. A. Lugiato, A. Gatti, and E. Brambilla, J.
Opt. B: Quantum Semiclassical Opt. \textbf{4}, S176 (2002), and
references therein.

\bibitem{Devaux00} F. Devaux and E. Lantz, Eur. Phys. J. D \textbf{8}, 117 (2000).

\bibitem{Jost98} B. M. Jost, A. V. Sergienko, A. F. Abouraddy,
B. E. A. Saleh, and M. C. Teich, Opt. Express \textbf{81}, 3
(1998).

\bibitem{Oemrawsingh02} S. S. R. Oemrawsingh, W. J. van Drunen, E.
R. Eliel, and J. P. Woerdman, J. Opt. Soc. Am. B \textbf{19}, 2391
(2002).

\bibitem{Aytur90} O. Aytur and P. Kumar, Phys. Rev. Lett. \textbf{65}, 1551
(1990).

\bibitem{Marable02} M. L. Marable, S.-K. Choi, and P. Kumar, Opt.
Express \textbf{2}, 84 (1998).

\bibitem{Brambilla04} E. Brambilla, A. Gatti, M. Bache, and L. A.
Lugiato, Phys. Rev. A \textbf{69}, 023802 (2004).

\bibitem{Gatti99} A. Gatti, E. Brambilla, and L. A. Lugiato, Phys.
Rev. Lett. \textbf{83}, 1763 (1999).

\bibitem{Yariv89} A. Yariv, {\it Quantum electronics} (John Wiley and Sons, New
York, 1989).

\bibitem{Berzanskis99} A. Berzanskis, W. Chinaglia, L. A. Lugiato,
K.-H. Feller, and P. Di Trapani, Phys. Rev. A \textbf{60}, 1626
(1999).

\bibitem{Rubin96} M. H. Rubin, Phys. Rev. A \textbf{54}, 5349 (1996).

\bibitem{Janesick01} J.R. Janesick, {\it Scientific Charge-Coupled Devices}
(SPIE Press Bellingham, Washington, 2001), pp. 204-205; see also
http://www.roperscientific.de/theory.html.

\bibitem{Jiang03} Y.-K. Jiang, O. Jedrkiewicz, S. Minardi, P. Di Trapani,
A. Mosset, E. Lantz, and F. Devaux, Eur. Phys. J. D \textbf{22},
521 (2003).

\bibitem{Koch} K. Koch, E. Cheung, G. T. Moore, S. H. Chakmakjian,
and J. M. Liu, IEEE J. of quant. electron. \textbf{31}, 769
(1995).

\bibitem{Mosset04} A. Mosset, F. Devaux, G. Fanjoux, and E. Lantz,
Eur. Phys. J. D \textbf{28}, 447 (2004).

\bibitem{Akhmanov} S. A. Akhmanov, V. A. Vysloukh, A. S. Chirkin,
\emph{Optics of Femtosecond Laser Pulses} (American Institute of
Physics, New York, 1992), p.151.

\bibitem{Ditrapani98} P. Di Trapani, G. Valiulis, W. Chinaglia and A. Andreoni,
Phys. Rev. Lett. \textbf{80}, 265 (1998).

\end{references}
\end{document}